\documentclass[10]{dles13}
\usepackage{graphicx,graphics,epsfig}
\usepackage{url}
\usepackage[square,sort,comma,numbers]{natbib}
\title{The effect of wing-tip vortices on the flow around a NACA0012 wing}
\author{{\bfseries  S. Toosi$^{1}$, A. Peplinski$^{1,2}$, P. Schlatter$^{1,2}$, R. Vinuesa$^{1,2}$ }, \\
  $^{1}$ {FLOW, Engineering Mechanics, KTH Royal Institute of Technology, Sweden} \\
  $^{2}$ {Swedish e-Science Research Centre (SeRC), Stockholm, Sweden} \\
\underline{siavasht@kth.se} }

\begin{document}
\maketitle
%
%%%%%%%%%%%%%%%%%%%%%%%%%%%%%%%%%%%%%%%%%%%%%%%%%%%%%%%%%%%%%%%%%%%%%%%%
% Main document - section example
%%%%%%%%%%%%%%%%%%%%%%%%%%%%%%%%%%%%%%%%%%%%%%%%%%%%%%%%%%%%%%%%%%%%%%%%
%
%
%%%%%%%%%%%%%%%%%%%%%%%%
% section example
%%%%%%%%%%%%%%%%%%%%%%%%
%
\section{Introduction}
% The two basic contributions to the total drag of an airplane are~\cite{phak} the parasite drag (including friction, form, and interference drags) and 
% the induced drag (vortex drag, or drag due to lift), which is caused by lift generation in finite-span wings and 
% the consequent presence of wing-tip vortices.
There are two basic contributions to the total drag of an airplane~\cite{phak}:
(i) parasite drag, which consists of friction drag, form drag, and interference drag, and (ii) induced drag (vortex drag, or drag due to lift), which is caused by lift generation in finite-span wings and 
the consequent presence of wing-tip vortices.
The two types of drags have close to equal contributions to the total drag in cruise conditions, 
while the induced drag is the dominant source during the take-off, climb and landing phases of flight~\cite{phak}.

The goal of the present work is to perform a systematic study of the formation of wing-tip vortices 
and their interaction with and impact on the surrounding flow in more details. 
Of particular interest is the interaction of these vortices with wall turbulence and the turbulent wake.
This is done by considering two wing geometries, i.e., infinite-span (periodic) and three-dimensional (wing-tip) wings,
at three different angles of attack: $\alpha=0^o,\,5^o,\,10^o$.

\section{Numerical method}

To ensure of the accuracy of the results,
we perform 
% a combination of direct numerical simulation (DNS) and 
high-resolution large-eddy simulations (LES), 
where only the smallest scales (e.g., $\leq 6\eta$ in the wake, where $\eta$ is the Kolmogorov length scale) 
are accounted for by the subgrid scale (SGS) model.
% where nearly all scales of the flow, including most of the dissipative ones, are resolved by the grid,
% and only the smallest scales (e.g., $\leq 6\eta$ in the wake, where $\eta$ is the Kolmogorov length scale) 
% are accounted for by the subgrid scale (SGS) model.
% 
% 
% 
Simulations are performed by the high-order incompressible Navier$-$Stokes solver Nek5000~\cite{nek5000}
with added adaptive mesh refinement (AMR) capabilities developed at KTH~\cite{nekAMR}.
% The standard version of Nek5000 only works with hexahedral elements of conforming topology (i.e., no ``hanging nodes'' are allowed). 
The AMR version adds the capability of handling non-conforming hexahedral elements with hanging nodes, and so, adds an $h$-refinement capability where each element can be refined individually. 
% (note that the polynomial order $p$ remains a global constant for all elements). 
Solution continuity at non-conforming interfaces is ensured by interpolating from the ``coarse'' side onto the fine side. 
% There are some modifications to the pressure solver and preconditioner~\cite{??}, stiffness matrix and its direct summation operations~\cite{??}, and similar, but the majority of the code is identical to the standard version.

The velocity field is expanded by a polynomial of order $p=7$ on the Gauss$-$Lobatto$-$Legendre (GLL) points 
($N=p+1=8$ GLL points in each direction), while the pressure is represented on $p-1$ Gauss$-$Legendre (GL) points following the $P_N-P_{N-2}$ formulation~\citep{nek5000}.
The nonlinear convective term is overintegrated 
% on a grid with $3N/2$ GL points in each direction, 
to avoid (or reduce) aliasing errors. 
Time stepping is performed by an implicit third-order backward-differentiation scheme for the viscous terms and an explicit third-order extrapolation for the nonlinear terms.
A high-pass-filter relaxation term~\citep{schlatter:2004} is added to the right-hand side of the equations, providing numerical stability and acting as a SGS dissipation.

Wings are located such that their mid-point along the chord line coincides with the origin of the coordinate system and have a no-slip no-penetration boundary condition.
Here $x$, $y$ and $z$ denote the streamwise, wall-normal and spanwise directions, respectively.
The computational domain has a rectangular cross-section in the $xy$-plane that extends $20c$ upstream, $30c$ downstream, and $20c$ in positive and negative $y$ directions.
Different angles of attack are achieved by rotating the wing along the $z$ axis around its center. %without changing the inflow boundary condition. 
This specific design is to allow for the use of the ``outflow-normal'' boundary condition~\citep{nek5000} on the $y$-normal boundaries; this allows for a non-zero $y$ component of velocity.
The three-dimensional (3D) wing domain extends $20c$ from the wing root (located at $z=0$) in the spanwise direction and has an ``outflow-normal'' boundary condition at $z=20c$. A symmetry boundary condition is used at the $z=0$ plane.
Boundary layers are tripped on both the suction and pressure sides of the airfoils for all six cases. 
% The tripping force is described in~\citep{??} with a small modification of multiplying by a Gaussian kernel to restrict the forcing to a few elements near the tripping lines.

\begin{figure}[t!]
\centerline{
 \includegraphics[width=0.95\linewidth]{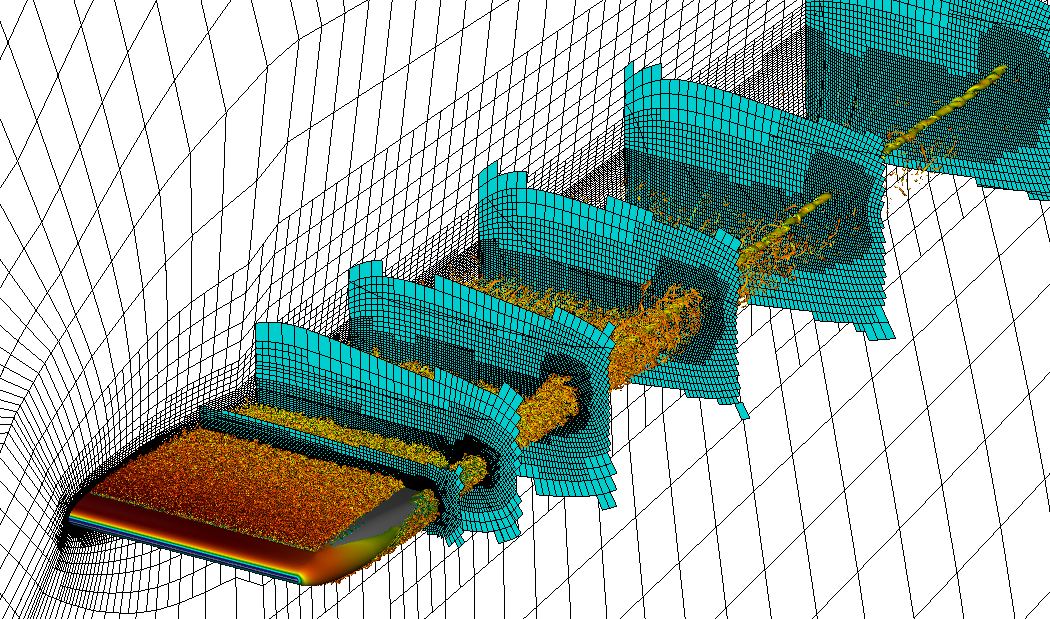}} 
\caption{Spectral-element grid for case RWT-5 with 3.1 million spectral elements. The grid is generated using the $h$-adaptation capabilities of the AMR version of Nek5000.
We show instantaneous vortical structures represented by $\lambda_2=-100$ isosurface colored by streamwise velocity ranging from (blue) low to (red) high.}
\label{fig:rwt5}
\end{figure}

% The production grids used in this study are generated by adapting (i.e., refining) an initial grid using the solution-based error indicator introduced by Mavripilis~\cite{??} for turbulent flows. 
The production grids are generated by iterative refinement of the initial grids, where at each iteration the elements with the highest contribution to solution error~\cite{fidkowski:2011}, based on solution on that grid, are selected for refinement. The error indicator of Mavriplis~\cite{mavriplis_90} is used for this purpose.
The convergence process is accelerated by some manual input from the user.
% for instance, 
% {\it e.g.}, by manually marking the wall elements for refinement (done in the first four refinement iterations), a fact that also compensates for the lack of the adjoint fields when selecting the elements. 
The adaptation process is terminated based on the resolution criteria available in the literature~\cite{verification_wing}.
Table~\ref{table:grids} summarizes some information about the production grids used in this work, and
Figure~\ref{fig:rwt5} shows the spectral elements of grid RWT-5 from Table~\ref{table:grids}.

\begin{table}[t!]
  \begin{center}
\def~{\hphantom{0}}
  \begin{tabular}{l | cc | c}
      Grid   & $N_{\rm GLL}$ & $N_{\rm grid}$ 
      & $(\Delta x^+_{\rm mean},y^+_1,\Delta z^+_{\rm mean})$  \\ \hline
      P-0    & $376\times10^6$  & $249\times10^6$ & (10.3,0.72,8.7)  \\
      P-5    & $381\times10^6$  & $252\times10^6$ & (10.5,0.73,8.5)  \\
      P-10   & $435\times10^6$  & $288\times10^6$ & (12,0.8,9)  \\
      RWT-0  & $950\times10^6$  & $630\times10^6$ & (10.3,0.7,5.5)  \\
      RWT-5  & $1.58\times10^9$ & $1.05\times10^9$ & (10.5,0.75,5.7)  \\
      RWT-10 & $2.16\times10^9$    & $1.43\times10^9$ & (12,0.8,6)  \\
      \hline
  \end{tabular}
  \caption{\label{table:grids}
  Summary of the production grids used in this study and their near-wall resolution.
  $N_{\rm GLL}$ shows the number of GLL points, while $N_{\rm grid}$ denotes the number of independent grid points.
  $\Delta_{\rm mean}=\delta/p$ is the average resolution of the element, where $\delta$ and $p$ are the element size and polynomial order.
  $y_1$ shows the distance of the second GLL point to the wall.
  Wall resolutions are normalized by the viscous length $\delta_{\nu}=\nu/\sqrt{\tau_w/\rho}$.
  The P-* grids denote periodic cases, whereas RWT-* stands for rounded wing-tip cases.
  The number next to each grid shows the angle of attack, $\alpha$.
%   The resolution of P-10 and all reported values for RWT-10 are estimates.
%  $N_{\rm GLL}$ shows the number of GLL points, $N_{\rm GLL}=N_{\rm el}\times(N+1)^3$, where $N_{\rm el}$ is the number of spectral elements, and $N_{\rm dof}$ shows the number of degrees of freedom, $N_{\rm dof} \approx N_{\rm el}\times N^3$.
%  $\Delta_{\rm mean}$ is the element size $h$ divided by polynomial order $p$, whereas $\delta{n_1}^+$ shows the distance from the wall of the first GLL point.
%  Wall resolutions are reported in wall coordinates ($s$ and $n$ denoting streamwise and wall-normal directions, respectively) and
%  normalized by the friction length, $\delta_\nu=\nu\sqrt{\rho/\tau_{\rm wall}}$, and
%  are reported at $0.7c$ from the leading edge, and $0.3c$ from the root in the RWT grids, at the element next to the wall. 
%  Wake resolutions are normalized by the approximate shear layer thickness, $\delta_{\rm shear}$, and
%  are reported at $2c$ downstream of the trailing edge, and $0.3c$ from the symmetry plane in the RWT grids, at the inflection point of the mean velocity.
  }
  \end{center}
\end{table}

\section{Results and discussion}
Some preliminary results are presented in this section.
These results are from the production grids of Table~\ref{table:grids} but have not been averaged for a sufficiently long period of time yet. Nevertheless, the observed phenomena are not expected to change for longer integration times.

\begin{figure}[t!]
\centerline{
\includegraphics[width=0.5\linewidth,clip=true,trim=0mm 0mm 33mm 10mm]{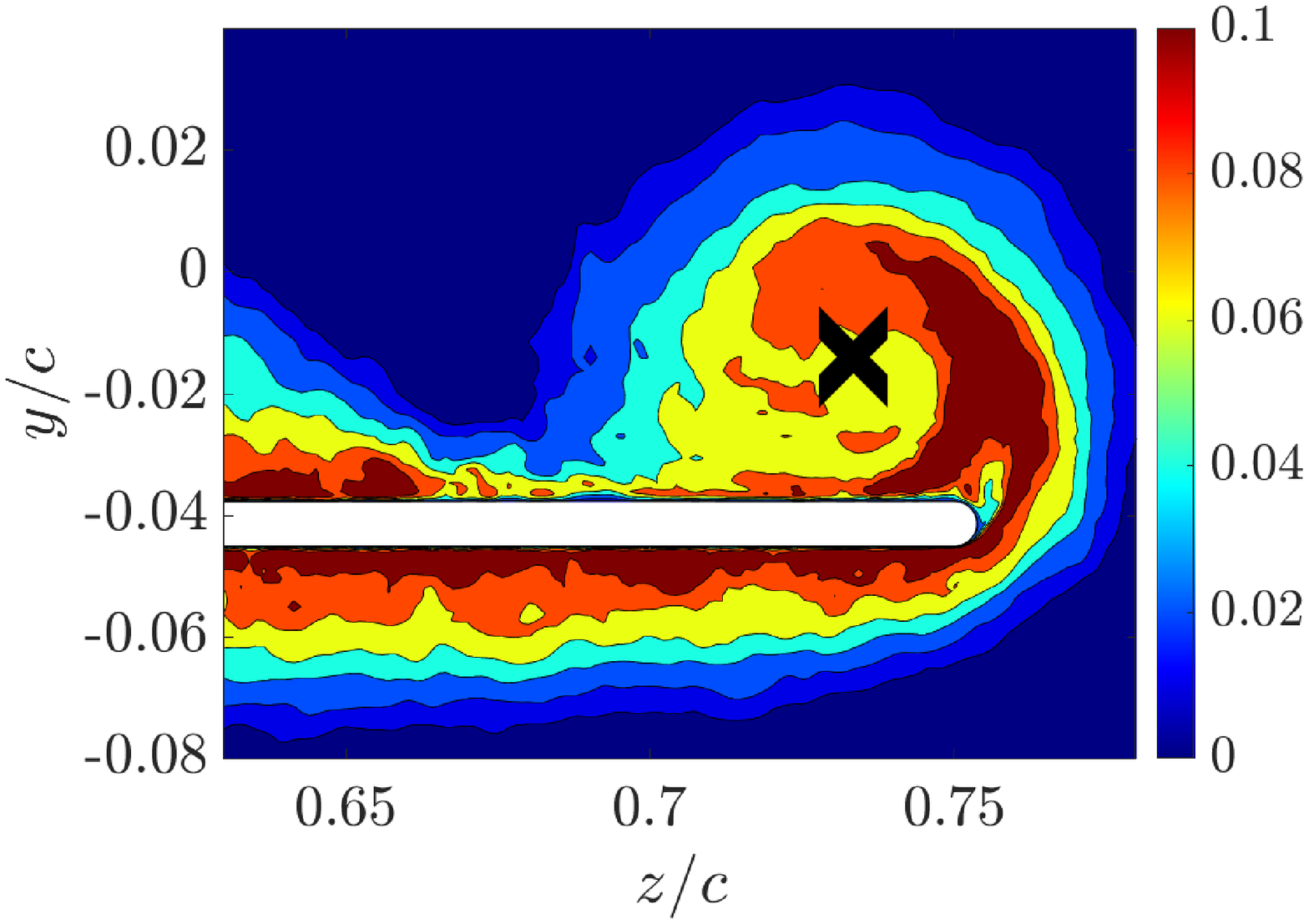} 
\includegraphics[width=0.5\linewidth,clip=true,trim=30mm 0mm 3mm 10mm]{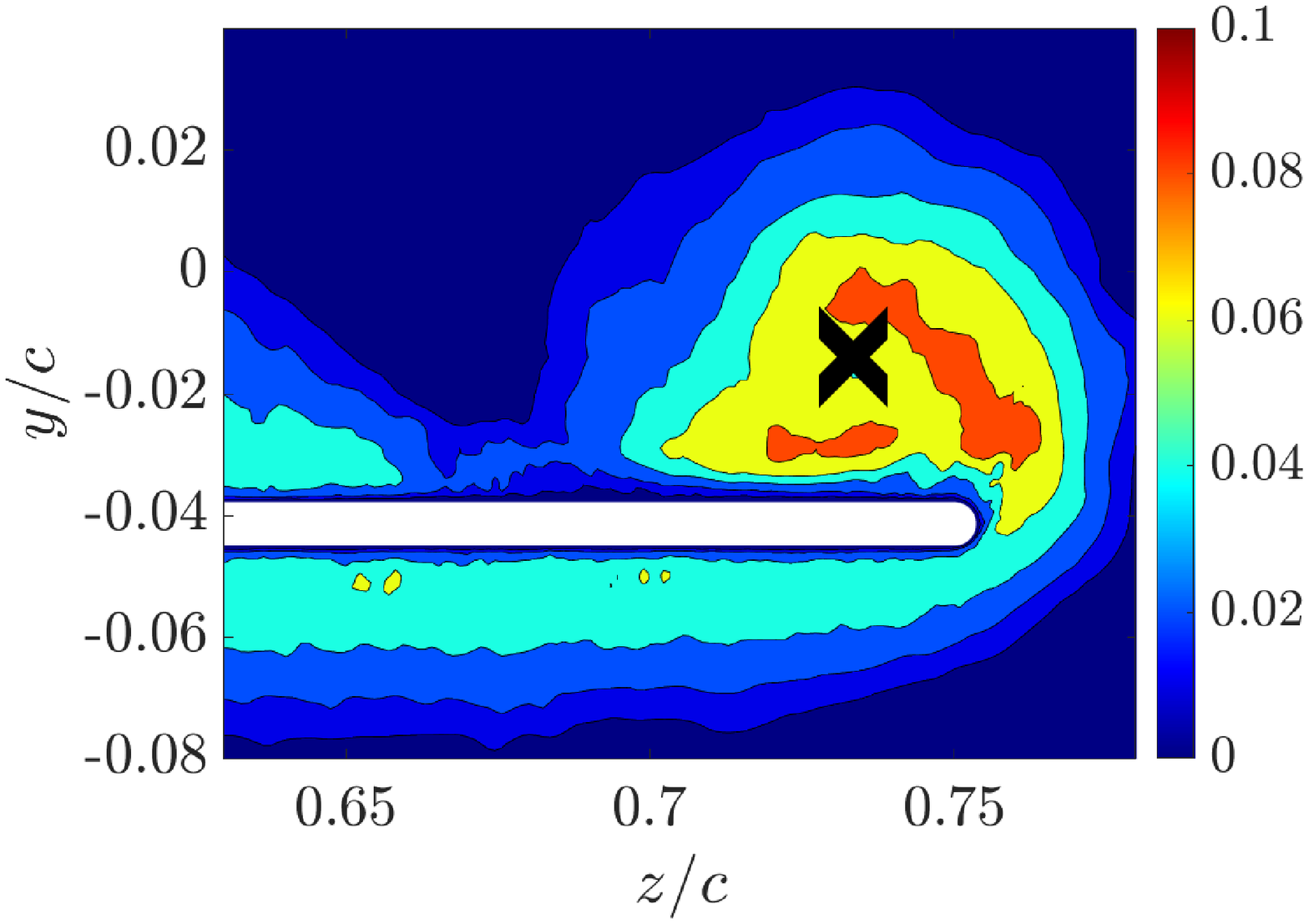} }
\caption{The rms velocities in the streamwise (left) and wall-normal (right) directions for the RWT-5 grid (Table~\ref{table:grids}) near the trailing edge.
The black X shows the location of the core of the wing-tip vortex.
The apparent noise is due to insufficient time averaging. 
}
\label{fig:R11_R22}
\end{figure}

Figure~\ref{fig:R11_R22} shows the root-mean-square (rms) velocities in the streamwise and wall-normal directions near the trailing edge of the airfoil on a $yz$-plane. It can be observed that the two components are non-zero and very close in magnitude to one another. We also notice that the strong rotation and pressure gradient imposed by the vortex tends to relaminarize the turbulent boundary layer on the suction side, as seen for instance at spanwise locations $0.66 \leq z/c \leq 0.7$.
This relaminarization can also be seen in the instantaneous vortical structures shown in Figure~\ref{fig:rwt5}.

\begin{figure}[t!]
\centerline{
\includegraphics[width=0.5\linewidth,clip=true,trim=0mm 0mm 32mm 10mm]{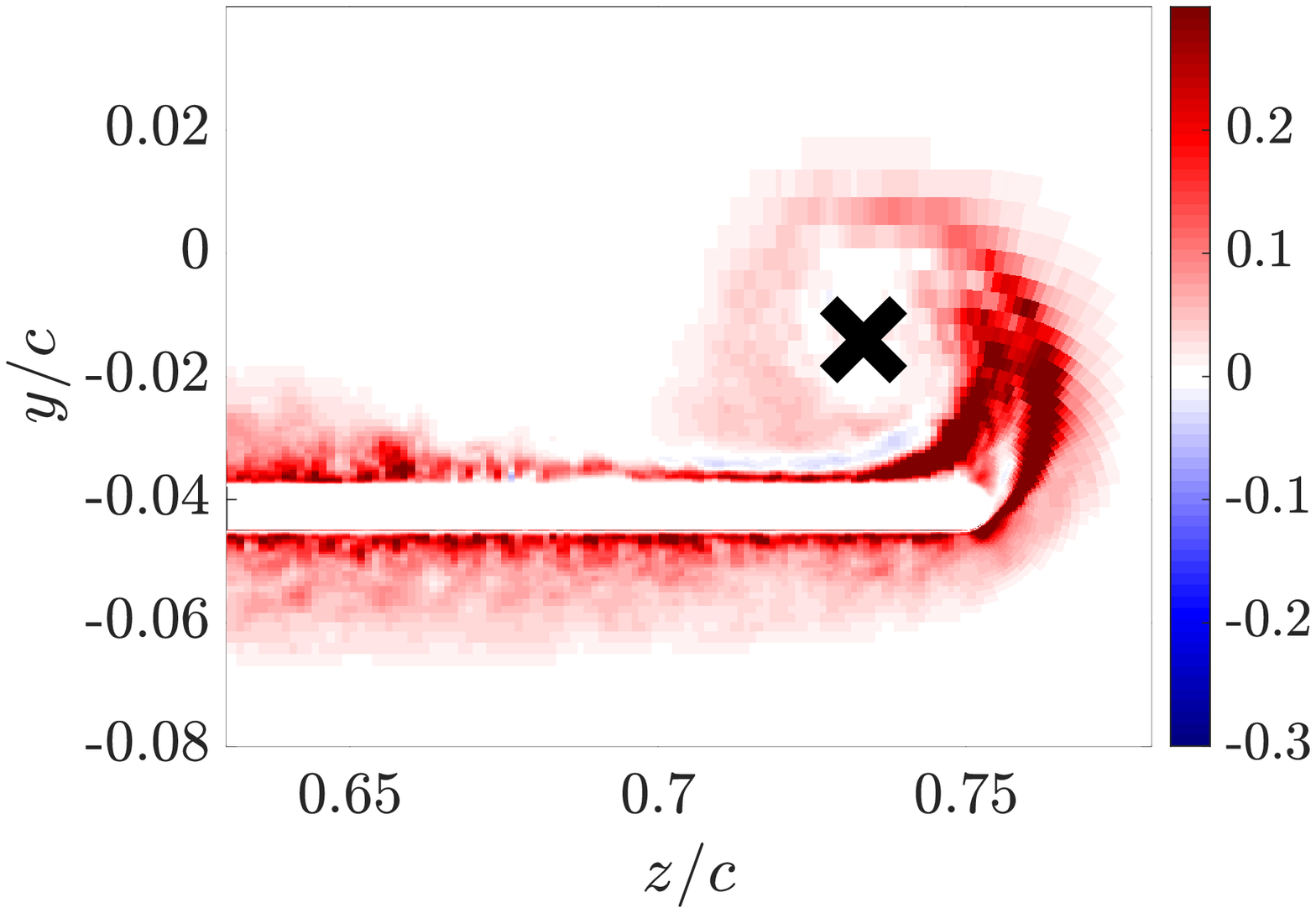} 
\includegraphics[width=0.5\linewidth,clip=true,trim=30mm 0mm 2mm 10mm]{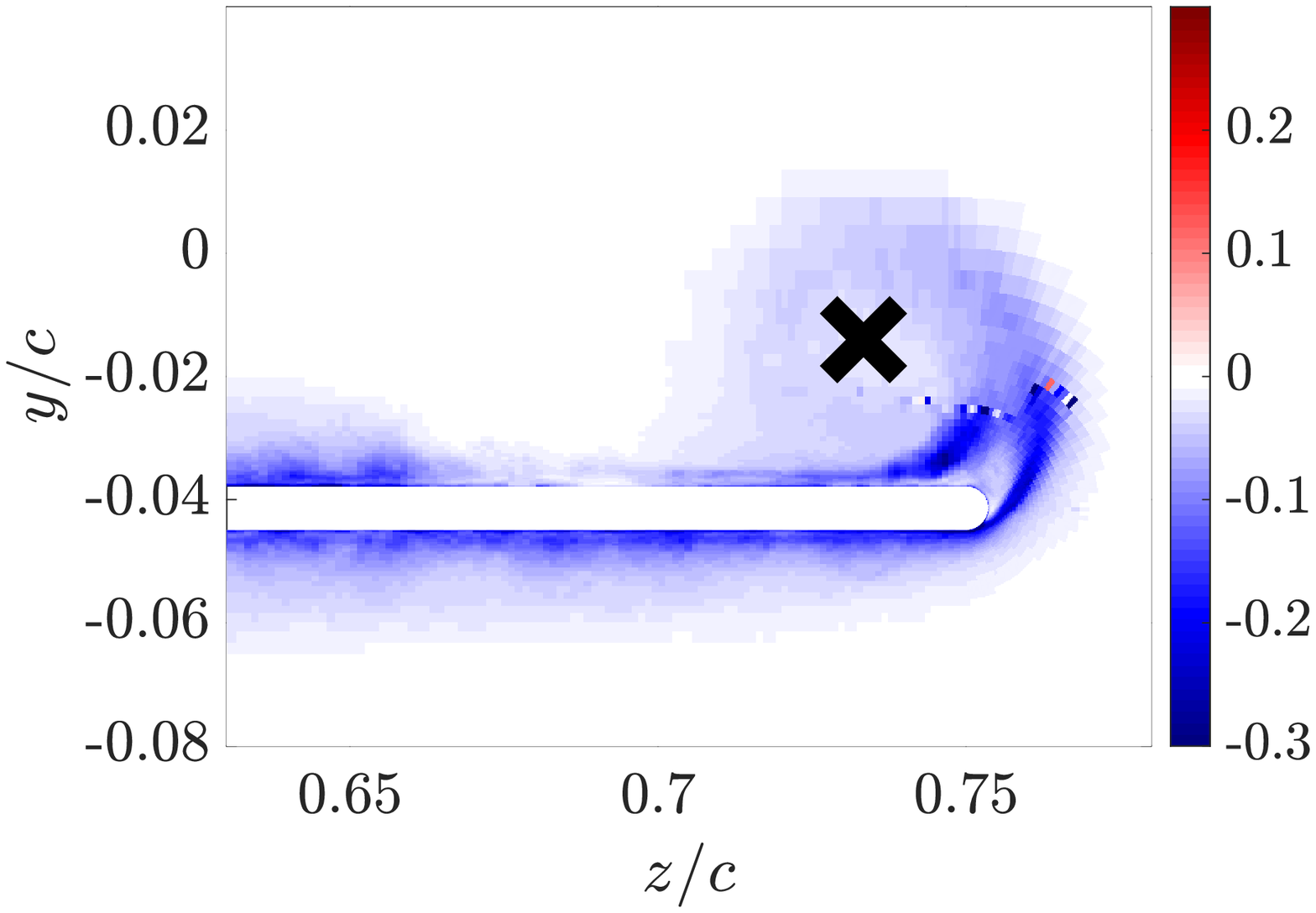}
 } 
\caption{Turbulent kinetic energy (TKE) production (left) and dissipation (right) at the same location as Figure~\ref{fig:R11_R22}.
The black X shows the location of the core of the wing-tip vortex.
The apparent noise is due to insufficient time averaging. 
}
\label{fig:TKE}
\end{figure}

Another interesting observation from Figure~\ref{fig:R11_R22} is the largermagnitude of velocity fluctuations in the vortex region outside the core to the top of the wing-tip, as well as the lower fluctuations closer to the wing-tip surface.
% (but presumably far enough not to be directly impacted by the no-slip, no-penetration condition at the wall).
A very similar pattern to these fluctuations is observed in the turbulent kinetic energy (TKE) production and dissipation terms shown in Figure~\ref{fig:TKE}, where we note a significant production region in the high-fluctuation area and very small production in the low-fluctuation regions.
This can be most simply explained by the functional form of the production term,
% a Reynolds stress component multiplied by a velocity gradient, 
which has higher values where both the Reynolds stresses (i.e., velocity fluctuations) and velocity gradients (near the vortex core) are high.
% Note that this observation is somewhat in contrast with the earlier belief 
% % found in the literature~\citet{?,?} 
% that most of the turbulent fluctuations in the vortex core are transported from the TBL turbulence.
Also notice that the TKE dissipation rate does not completely match its production in shape or magnitude, suggesting a highly non-local turbulence with significant transport phenomena.

% \section{Summary and conclusions}
\section{Conclusions and outlook}

This work leverages the AMR version of Nek5000 to perform several high-resolution large-eddy simulations of the flow around periodic and 3D NACA0012 wings to
investigates the impact of wing-tip vortices on the flow. 

Preliminary results suggest significant TKE production as a result of wing-tip formation, as well as a non-local turbulence characterized by the local imbalance between TKE production and dissipation. 

The final work 
to be presented in the conference will be based on longer integration times with converged statistics, and will also include spectral analysis.
% expands on the brief investigation of this flow presented here while benefiting from longer integration times.

\begin{References}

\bibliographystyle{ieeetr}

\bibitem[Federal Aviation Administration, 2016] {phak} 
Federal Aviation Administration :
Pilot’s Handbook of Aeronautical Knowledge, 
{\em United States Department of Transportation, Federal Aviation Administration, Airman Testing Standards Branch, Oklahoma City, United States}, (2016).

\bibitem{nek5000} Fischer P.F., Lottes J.W., Kerkemeier S.G. : NEK5000: Open source spectral element CFD solver. {\em Available at: \url{http://nek5000.mcs.anl.gov}}, (2008).

% \bibitem{nekAMR} Peplinski A., Offermans N., Fischer P. F., Schlatter P. : Non-conforming elements in Nek5000: pressure preconditioning and parallel performance, {\em Spectral and High Order Methods for Partial Differential Equations ICOSAHOM 2018}, 599--609 (2018).

\bibitem{nekAMR} Offermans N., Peplinski A., Marin O., Schlatter P. : Adaptive mesh refinement for steady flows in Nek5000, {\em Computer \& Fluids}, \textbf{197}, 104352 (2020).

\bibitem{mavriplis_90} Mavriplis C. : A posteriori error estimators for adaptive spectral element techniques, {\em In: Wesseling, P. (ed.) Proceedings of the Eighth GAMM-Conference on Numerical Methods in Fluid Mechanics. Notes on Numerical Fluid Mechanics}, 333--342 (1990).

\bibitem{verification_wing} Vinuesa R., Negi P.S., Atzori M., Hanifi A., Henningson D.S., Schlatter P. : Turbulent boundary layers around wing sections up to $Re_c=1,000,000$. {\em International Journal of Heat and Fluid Flow}, \textbf{72}, 86--99 (2018).

\bibitem{fidkowski:2011} Fidkowski, K. J., Darmofal, D. L. : Review of Output-Based Error Estimation and Mesh Adaptation in Computational Fluid Dynamics, {\em AIAA Journal}, \textbf{49}, 673--694 (2011).

\bibitem{schlatter:2004} Schlatter, P., Stolz, S., Kleiser, L. : LES of transitional flows using the approximate deconvolution model, {\em International Journal of Heat and Fluid Flow}, \textbf{25}, 549--558 (2004).
 
% \bibitem[Bruno et~al., 2014] {Bruno2014} Bruno, L., Salvetti, M.V. and Ricciardelli, F. : Benchmark on the aerodynamics of a rectangular 5:1 cylinder: and overview after the first four years of activity, {\em J. Wind Eng. Ind. Aerod.}, \textbf{126}, 87--106 (2014).
% \bibitem[Salvetti and Bruno, 2013] {Salvetti2014} Salvetti, M.V. and Bruno, L. : Reliability of LES simulations in the context of a benchmark on the aerodynamics of a rectangular 5:1 cylinder, {\em Proc. of Direct and Large-Eddy Simulations 9, April 3-5, Dresden, Germany}, (2013).

\end{References}

\end{document}